\documentclass{PoS}

\usepackage{graphicx}
\usepackage{epsfig}

\title{Gluon Mass in Landau Gauge QCD}

\ShortTitle{Gluon Mass in Landau Gauge QCD}

\author{\speaker{P. Bicudo}\\
            CFTP, Departamento de F\'{i}sica, Instituto Superior T\'ecnico, Av. Rovisco Pais, 1049-001 Lisboa, Portugal \\
        E-mail: \email{bicudo@ist.utl.pt}}

\author{O. Oliveira\\
        Departamento de F\'{\i}sica, Universidade de Coimbra, 3004-516 Coimbra, Portugal\\
        E-mail: \email{orlando@teor.fis.uc.pt}}

\abstract{The interpretation of the Landau gauge lattice gluon propagator as a massive type bosonic propagator is investigated for
i) an infrared constant gluon mass; ii) an ultraviolet constant gluon mass;  iii) a momentum dependent mass. 
We find that the infrared data can be associated with a massive propagator with a constant gluon mass of 651(12) MeV, but
the ultraviolet lattice data is not compatible this type of propagator.
The scenario of a momentum dependent gluon mass gives a decreasing mass with the momentum, starting from a value of 
$\sim 630$ MeV in the infrared region and suggesting a $q^2 \ln q^2$ dependence for momenta above 1 GeV. 
}

\FullConference{The XXVIII International Symposium on Lattice Field Theory, Lattice2010\\
		June 14-19, 2010\\
		Villasimius, Italy}

\begin{document}

\section{Introduction and Motivation}

If at the classical level SU(3) Yang-Mills theory is conformal invariant, the corresponding quantum theory breaks this symmetry
via dimensional transmutation. Indeed, dimensional transmutation introduces a scale in QCD, $\Lambda_{QCD} \sim 300$ MeV,
which defines the typical energy for strong interactions. 
Despite the breaking of conformal invariance, at the level of the lagrangian a mass term is forbidden 
by gauge invariance. Moreover, within the perturbative solution of QCD in Landau gauge the gluon is a massless particle.
However, if one goes beyond perturbation theory and looks for nonperturbative solutions of theory, then
a dynamical mass for the gluon becomes possible \cite{Cornwall82}. This gluon mass is, typically, a function of the 
gluon momentum $M(q^2)$.

A non-vanishing gluon mass is welcome to regularize infrared divergences  and solve some problems related with unitarity. 
Diffractive phenomena \cite{Forshaw99} and inclusive radiative decays of $J / \psi$ and $\Upsilon$ \cite{Field02} 
suggest a massive gluon with a mass in the range 0.500 -- 1.2 GeV depending on how you define the mass.
Moreover, lattice simulations also suggest an infrared gluon hard mass of $\sim 600$ MeV \cite{Oliveira09} 
and an ultraviolet mass $M_g \sim 1.0$ GeV \cite{Leinweber99,Silva04}. 

A massive gluon is also welcome within the dual picture of the QCD vaccuum, where a Meissner effect due to 
chromomagnetic Abrikosov flux tubes introduces an effective gluon mass \cite{Nambu,Hooft,Mandelstam}.

From the point of view of the Dyson-Schwinger equations, the idea of a gluon mass which as a function of
the gluon momenta fits naturally within the so-called decoupling solution
\cite{Aguilar03,Aguilar08,Aguilar08b,Cornwall09,Fischer09}.
Indeed, the numerical solutions of the DSE give a  $M(q^2)$ which takes its largest value at zero momentum, 
where $M(0) \sim 600$ MeV, and vanishes for $q \gg \Lambda_{QCD}$, recovering, in this way, the usual perturbative propagator 
at high momentum. 

Lattice QCD simulations also provide support for a non-vanishing gluon mass -- see, for example,
 \cite{Leinweber99,Silva04,Oliveira09,Oliveira10} and references there in. 
The precise value for $M(q^2)$ depends on how the gluon propagator is modeled. 

\section{The Gluon Propagator and The Gluon Mass}

In the Landau gauge the momentum space gluon propagator is given by
\begin{equation}
  D^{ab}_{\mu\nu} (q^2) ~ = ~  \langle A^a_\mu(q) A^b_\nu (-q) \rangle 
                                       ~ = ~ \delta^{ab} \, \left( \delta_{\mu\nu} - \frac{q_\mu q_\nu}{q^2} \right) \, D(q^2) \, .
\end{equation}
As definition of the gluon mass we take
\begin{equation}
   D(q^2) = \frac{Z(q^2)}{q^2 + M^2 (q^2)} 
   \label{DZM}
\end{equation}
and, in the following, we will measure $Z(q^2)$ and
$M^2 (q^2)$ from the gluon data computed in SU(3) lattice QCD simulations at $\beta = 6.0$ for
various volumes. Details of the simulation can be found in \cite{Oliveira10}.

The gluon propagator used to compute $M^2(q^2)$ are renormalized at $\mu = 3$ GeV such that $D(\mu^2) = 1 / \mu^2$ -- see
 \cite{Oliveira10} for details.

The lattice spacing effects are removed/reduced applying, for each volume, a conic cut for momenta $q > 1$ GeV \cite{Leinweber99}.
To achieve a good description of the infrared region, for momenta smaller than 1 GeV all the data points coming from the various
simulations were considered. Further cuts were applied to reduce finite volume effects. The data from the smaller lattices
which deviate from our largest volume simulation $\sim (8.1 \mbox{ fm})^4$ were deleted from the data set. Then,
from the $64^4$ propagator only data with $q \ge 425$ MeV was considered, 
from $48^4$ only $q \ge 671$ MeV data was considered and from $32^4$ only data with $q \ge 848$ MeV was included.
However, the IR and UV cuts described are not able to remove all lattice spacing and volume effects. 
For each lattice and for the same $q^2$ coming from different $q_\mu$, if the different estimates of the propagator don't agree 
within one standard deviation, one of the points is removed. For example, for the $80^4$ lattice for momentum $q = 457$ MeV 
there are two estimates for the gluon propagator, $D(q^2) = 6.515(96)$ GeV$^{-2}$ and $D(q^2) =  6.382(64)$ GeV$^{-2}$, 
coming from different types of momenta.
The first value is clearly above all the data points and it was not considered in the combined data set.
In this way, the surviving points will produce a unique curve for $D(q^2)$.

\begin{figure}[t]
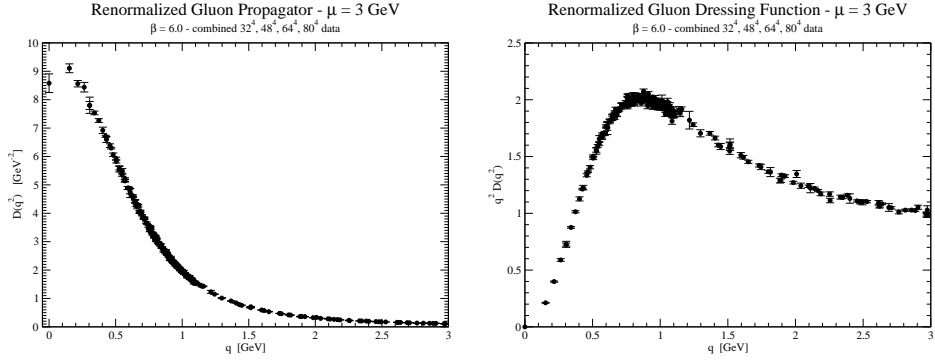

  \centering
   \includegraphics[scale=0.25]{figure2.eps}  \quad 
   \includegraphics[scale=0.25]{figure3.eps} 
   \caption{Renormalized gluon propagator $D(q^2)$ (left) and gluon dressing function $q^2 D(q^2)$ (right).}
   \label{fig_dress}
\end{figure}

The combined lattice data for gluon propagator, after all the cuts, is shown in figure \ref{fig_dress} together with
the dressing function $q^2 D(q^2)$.

\section{An Infrared Constant Mass \label{section_hadr_ir}}

\begin{figure}[t]
   \centering
   \includegraphics[scale=0.25]{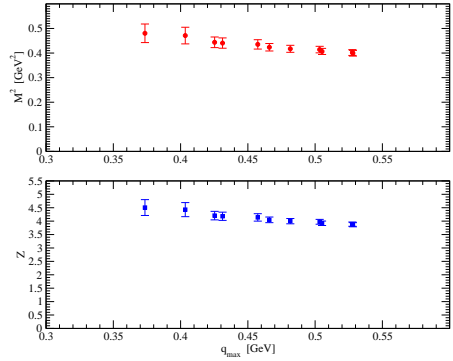}
     \caption{Fitting an infrared constant mass ($\chi^2/d.o.f. \le 1.8$).}
   \label{fig_hard_mass} 
\end{figure}

Let us start our discussion on the gluon mass considering the case of an infrared  constant gluon mass, i.e. assuming that the
gluon propagator is described by
\begin{equation}
   D(q^2) = \frac{Z}{q^2 + M^2} \, ,
   \label{hard_mass}
\end{equation}
where $Z$ and $M$ are constants,  in the momentum range $[0, q_{max}]$. 
The results of fitting the lattice data to (\ref{hard_mass}) are plotted in figure \ref{fig_hard_mass}.

Figure \ref{fig_hard_mass} shows a $M^2$ and $Z$ that are, within one standard deviation, stable against variation of
$q_{max}$. Furthermore, demanding that $\chi^2/d.o.f. < 1.8$, then the infrared propagator can be described by equation
(\ref{hard_mass}) for momenta up to $q \sim 530$ MeV. The fits reported here have a $1.26 \le \chi^2/d.o.f.< 1.8$.

From the above results one can conclude that, with the possible exception of the deep infrared region, 
$D(q^2)$ can be described by a massive type propagator with a constant mass in the low energy regime. 

\begin{figure}[t]
\vspace{0.2cm}
   \centering
   \includegraphics[scale=0.25]{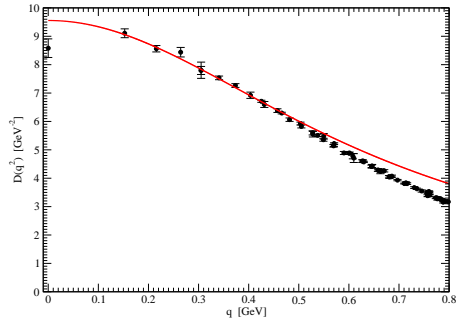} 
   \caption{Lattice gluon propagator and the infrared fit the smallest $\chi^2/d.o.f.$}
   \label{figfit_hard_mass}
\end{figure}

The fit with the lower $\chi^2/d.o.f.$ has $q_{max} = 466$ MeV, $Z =  4.05(10)$ and $M =  651(12)$ MeV.
The fit together with the lattice propagator are reported in figure \ref{figfit_hard_mass}. 

\section{An Ultraviolet Constant Mass}

If one applies the same reasoning to the ultraviolet (UV) region, i.e. for $ q > 2.5$ GeV, it turns out that
$M^2$ is not stable against variation of the range of momenta considered. In this sense one cannot define a gluon constant mass
to the ultraviolet region. This is not a surprise. Indeed to renormalize the lattice propagator one uses a perturbative inspired 
one-loop expression, which describes very well the data in the UV region -- see \cite{Oliveira10} for details.

The results discussed here for the UV are not in contradiction with thise of \cite{Leinweber99,Silva04},
where an ultraviolet gluon mass of $\sim 1 $ GeV was claimed. In \cite{Leinweber99,Silva04} an ultraviolet regulator was used and 
the full set of lattice data surviving the conic cut fitted to
\begin{equation}
   D(q^2 ) = Z \, \frac{ \Big[ \frac{1}{2} \ln (q^2 + M^2) (q^{-2} + M^{-2}) \Big]^{-\gamma} }{q^2 + M^2}\, ,
   \label{UV_reg}
\end{equation}   
where $M$ is the gluon mass. The difference is due to different definitions for the gluon mass.

In conclusion, in what concerns the ultraviolet region, the lattice gluon propagator is not described by a constant massive propagator.

\section{A Momentum Dependent Massive Gluon }

\begin{figure}[t]
   \centering
   \includegraphics[scale=0.25]{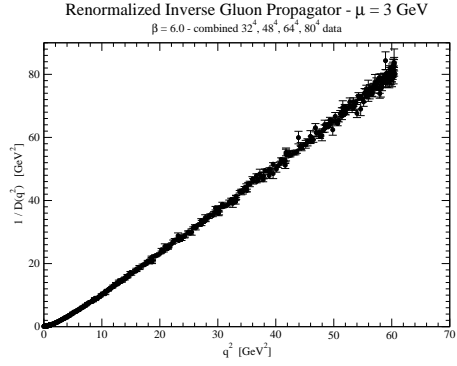} 
   \caption{Inverse gluon propagator - note the "almost" linear behaviour with $q^2$ of $1/D(q^2)$.}
   \label{fig_inv_prop}
\end{figure}

\begin{figure}[t]
\vspace{0.3cm}
   \centering
   \includegraphics[scale=0.25]{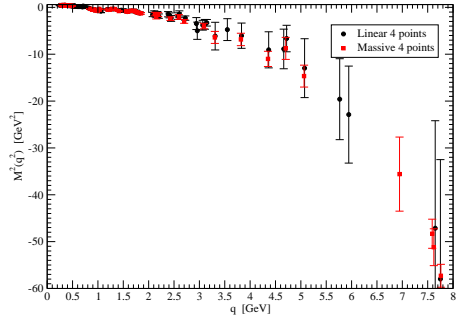} 
   \caption{Overall picture for running gluon mass - only data whose fit has $\chi^2/d.o.f. < 1.8$ and relative error less than 30\% 
                is included in the plot.}
   \label{fig_run_mass01}
\end{figure}

In this section we will assume a $D(q^2)$ given by equation (\ref{DZM}). 
The momentum functions $M(q^2)$ and $Z(q^2)$ will be referred as the running mass and running dressing function, respectively.

For the computation of $Z(q^2)$ and $M^2(q^2)$ two different methods called (i) linear 4 points and (ii)
massive 4 points in the figures, were considered. Furthermore, $Z(q^2)$ and $M^2(q^2)$ were investigated using both
$D(q^2)$, see figure \ref{fig_dress}, and $1/D(q^2)$, see figure \ref{fig_inv_prop}.
The interested reader can find the details on the procedure to extract the functions from the lattice data in \cite{Oliveira10}.
The fits with a $\chi^2/d.o.f. > 1.8$ were exclude from the analysis.

The running mass, computed with both methods, is shown in figure \ref{fig_run_mass01}. 
$M^2(q^2)$ is positive in the infrared region, decreases with $q$ and becomes negative around $q \sim 800$ MeV. 
Although $M^2(q^2)$ becomes negative, $q^2 + M^2(q^2)$ is always positive defined, i.e. 
the propagator has no poles for euclidean momenta.

For the infrared region, the running mass measured for the smallest momenta have $q = 249$ MeV (method i) 
and 273 MeV (method ii). The corresponding mass values are, respectively, $652(79)$ MeV and $669(88)$ MeV. 
We note the excellent agreement with the estimation of a hard infrared constant mass $651(12)$ MeV 
(see section \ref{section_hadr_ir}). 

Let us investigate a possible ansatz for $M^2(q^2)$. Our best fit occurs when the ultraviolet region and the infrared region are 
studied separately. Given that the statistical errors on $M^2(q^2)$ increase with $q$, it compromises the investigation of the 
ultraviolet behavior  -- see figure \ref{fig_run_mass01}. However, starting at a relatively low momenta, let us say around 1 GeV, 
one can test for the $q^2$ dependence of $M^2(q^2)$. 
Our best fit points towards a $M^2(q^2) \sim q^2 \ln q^2$ at high momentum and an infrared
dependence where $M^2(q^2) \sim q^2$. 

The outcome of the separate fits to the two momentum regions is
\begin{equation}
   M^2(q^2) = \left\{ \begin{array}{lll} 
           0.534(15) - 0.943(48) \, q^2 \, , &  & ~ \mbox{method (i)} \\
           0.578(37) - 1.239(64) \, q^2 \,  , &  & ~ \mbox{method (ii)}
           \end{array} \right.
           \label{m2_q2_IRfits}
\end{equation}  
with a $\chi^2/d.of.$ of 0.6 and 1.6, respectively, for the infrared region, i.e for $q < 1$ GeV, and
\begin{equation}
   M^2(q^2) = \left\{ \begin{array}{lll} 
           -0.349(22) - 0.1465(79) \, q^2 \ln q^2 \, , &  & ~ \mbox{method (i)}\\
           -0.288(33) - 0.2050(78) \, q^2 \ln q^2\,  , &  & ~ \mbox{method (ii)}
           \end{array} \right.
           \label{m2_q2_UVfits}
\end{equation}  
with a $\chi^2/d.of.$ of 0.5 and 1.8, respectively, for the UV region, i.e. for $q > 1$ GeV. 
In the above formula $M^2(q^2)$ and $q^2$ are given in GeV$^2$.

\begin{figure}[t]
   \centering
   \includegraphics[scale=0.25]{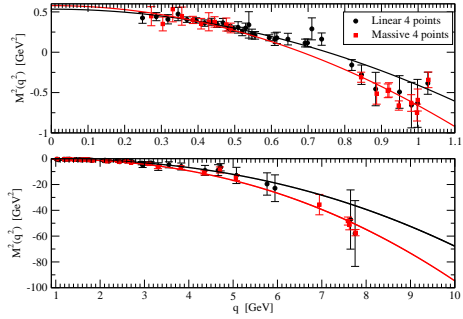}
   \caption{Running gluon mass $M^2(q^2)$ and the fits (full lines) to (5.1)  and  (5.2).}
   \label{fig_run_mass}
\end{figure}

In figure \ref{fig_run_mass} we plot the lattice data together with the fits to (\ref{m2_q2_IRfits}) and
(\ref{m2_q2_UVfits}). The fits to the infrared region, see equation (\ref{m2_q2_IRfits}), give an $M(0)$
731(11) MeV (method i)and 760(24) MeV (method ii), which is slightly larger than the constant infrared mass 
computed in section \ref{section_hadr_ir}. Anyway, the fits to the infrared region suggest a finite $D(0)$.
Moreover, the predicted $M(0)$ is associated with a $D(0) \sim 2$ GeV$^{-2}$, a value of 
the same order of magnitude as predicted by recent large volume lattice simulations \cite{Dudal10}.

\begin{figure}[t]
   \centering
   \includegraphics[scale=0.25]{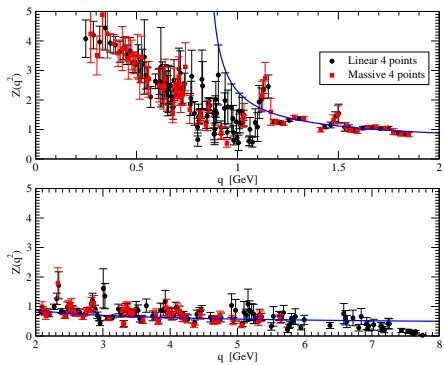}
   \caption{Running gluon dressing function $Z(q^2)$. Only data whose fit has a $\chi^2/d.o.f. < 1.8$ and relative error less than 30\% is 
                 included in  the plot. For the meaning of the full line, see the main text.}
   \label{fig_run_mass_z}
\end{figure}

The running gluon dressing function $Z(q^2)$ data is displayed in figure \ref{fig_run_mass_z} 
together with the dressing function the (full line) computed using the 1-loop perturbative expression
for $Z(q^2)$ rescaled by 0.7.

In the low momentum region, $Z(q^2)$ decreases from $q = 0$ up to 1 GeV.
However, when the zero momentum is approached from above, $Z(q^2)$ seems to saturate around $q = 400$ MeV. 
Unfortunately, the large statistical errors in $Z(q^2)$ for $q < 1$ GeV make it difficult to disentangle the functional dependence 
but, clearly, $Z(q^2)$ does not follows the perturbative behavior. 

The gluon dressing function $Z(q^2)$ is well described by the anstaz 
\begin{equation}
   Z(q^2) = \frac{Z_0}{\left[ A + \ln(q^2 + m^2_0) \right]^\gamma} \, ,
   \label{z_fit_func0}
\end{equation}    
where $\gamma = 13/22$ is the anomalous gluon dimension, over the full range of momenta. The fits
give $Z_0 = 1.048(75)$, $A = -0.43(21)$, $m^2_0 = 1.57(33)$ GeV$^2$ for a $\chi^2/d.o.f. = 1.8$ (method i) and
$Z_0 = 0.98(14)$, $A = -0.54(38)$, $m^2_0 = 1.76(68)$ GeV$^2$ with a $\chi^2/d.o.f. = 1.9$ (method ii).


\section*{Acknowledgments}

The authors acknowledge financial support from F.C.T. under project CERN/FP/83582/2008 and CERN/FP/109327/2009.
The authors thank A. Aguilar for helpful discussions. The authors thank P. J. Silva for the help with the gauge fixing for the $32^4$ lattice.


\end{document}